\documentstyle[12pt]{article}
\begin{document}
\newcommand{\nd}[1]{/\hspace{-0.5em} #1}
\begin{titlepage}
\begin{flushright}
SWAT-46 \\
hep-th/9410140 \\
October 1994  \\
\end{flushright}
\vspace{.6in}
\begin{centering}
{\Large {\bf  Quantum scattering of charged solitons in the complex}\\
\vspace{0.5cm}
{\bf sine-Gordon model}} \\
\vspace{.8in}
Nicholas Dorey and Timothy J. Hollowood \\
\vspace{.1in}
Department of Physics, University of Wales Swansea \\
Singleton Park, Swansea, SA2 8PP, UK \\
\vspace{1.4in}
{\bf Abstract} \\
\vspace{.05in}
\end{centering}
{\small The scattering of charged solitons in the
complex sine-Gordon field theory is investigated.
An exact factorizable S-matrix for the theory is proposed
when the renormalized coupling constant takes the values
$\lambda^{2}_{R}=4\pi/k$ for any integer $k>1$: the minimal
S-matrix associated with the Lie algebra
$a_{k-1}$. It is shown that the
proposed S-matrix reproduces the leading semiclassical behaviour
of all amplitudes in the theory and is the minimal S-matrix which is
consistent with the semiclassical spectrum of the model. The results
are completely consistent with the description of the complex
sine-Gordon theory as the SU$(2)/{\rm U}(1)$ coset model at level $k$
perturbed by its first thermal operator.}
\end{titlepage}
\section{Introduction}
\paragraph{}
The idea that extended particles in quantum field theory
can be associated with soliton solutions of the corresponding classical
field equations dates back to the early work of Skyrme \cite{sky}.
The sine-Gordon (SG) field theory is an exactly solvable model which
provides an important paradigm for this idea. However, the sine-Gordon kink
lacks a key feature which is present in more realistic four-dimensional
theories: it has no internal degrees of freedom.
All solitons, including the SG kink, have a translational degree of freedom
which corresponds to the position of the centre of mass, but the soliton
solutions of a field theory which has an unbroken
global symmetry will also
have internal collective coordinates. When the
theory
is quantized these coordinates give rise to a tower of
massive particles which carry the corresponding conserved charge.
One example of this phenomenon
occurs in the Skyrme model, where the space of static
soliton solutions is parametrized by an
SU$(2)$ collective coordinate. The corresponding particles have angular
momentum and isospin quantum numbers and are identified with the baryons of
large-$N_{c}$ QCD \cite{adk}.
Another example is the 't Hooft-Polyakov monopole in
non-abelian gauge theory \cite{tho,pol}.
The monopole has an internal U$(1)$ degree of
freedom which gives rise to dyons; particles which
carry both electric and magnetic charge \cite{jul}.
\paragraph{}
In both the examples mentioned above, the spectrum
and interactions of these particles can generally
only be found at weak coupling and so it would be useful
to have a simple model with internal symmetry which was also
exactly solvable in the sense of the SG theory.
In this paper we will
study such a model and propose an exact solution under certain
conditions. The complex sine-Gordon (CSG)
theory \cite{poh,get} is defined by the following Lagrangian in
two spacetime dimensions which has a global U$(1)$ symmetry:
\begin{equation}
{\cal L}=\frac{|\partial_{\mu}\psi|^{2}}{1-\lambda^{2}|\psi|^{2}}
-m^2|\psi|^{2},
\label{lag}
\end{equation}
where $\psi=\psi_{1}+i\psi_{2}$ is a complex scalar field. Like the
more familiar SG equation, the classical field theory which follows
from (\ref{lag}) has both single soliton solutions and multi-soliton
scattering solutions which can be constructed analytically
\cite{get,dev2}. In the classical
theory, the solitons exhibit completely elastic scattering, the only
effect of which is a time-delay relative to free motion. By analogy with the
SG case one might hope that the corresponding quantum theory would also be
completely integrable and would therefore have a factorizable S-matrix.
De Vega and Maillet \cite{dev1} calculated the scattering amplitude for the
elementary particles of the theory (which we will call mesons)
in ordinary perturbation theory and found
that the S-matrix is indeed factorizable at tree level. At the one-loop level
however, the factorizability of the theory breaks down, and can only be
regained if a local counter-term is added to the bare
Lagrangian. It is not known if this procedure can be carried out to
all orders to yield an exactly factorizable S-matrix for the mesons.
In this paper we will be led to the surprising result that the theory
only admits a factorizable S-matrix for particular values of the
renormalized coupling; namely $\lambda^{2}_{R}=4\pi/k$ where $k$ is
an integer greater than 1. For these values we will propose an exact S-matrix
for the CSG theory which is consistent both with the general
requirements of completely elastic scattering theory in two dimensions
\cite{zam} and with the semiclassical limit of the model. The
S-matrix in question is the minimal S-matrix associated
with the Lie algebra $a_{k-1}$ and in the present case the absence of
CDD ambiguities follows immediately from our semiclassical results.
\paragraph{}
Recently an alternative view of the theory has emerged
based on the work of Bakas \cite{bakas} (see also \cite{qhan}),
who showed at the
classical level that the CSG Lagrangian with coupling
constant $\lambda^{2}=4\pi/k$ is obtained as a particular gauge fixing
of a perturbation (by the first thermal operator) of the
$SU(2)/U(1)$ coset model at level $k$, realized as a gauged WZW model.
The coset description resolves the apparent
pathology of the model signaled by
the singularity at $|\psi|=1/\lambda$ in the Lagrangian, which is
nothing more than a bad choice of coordinates on the gauge-slice, and
provides a definition of the path-integral measure.
The results of our investigation are completely
consistent with this coset description of the model. In particular, the
restriction of $k$ to integer values which emerges in the CSG theory by
demanding an S-matrix which is consistent with the semiclassical
limit, has a natural explanation: the coset model is only
well-defined quantum mechanically if the level is an integer greater
than one. Also, finite counter-terms like those described above which are
necessary to maintain the factorizability of the S-matrix
arise naturally in the coset
model as $1/k$ corrections to the effective action derived from the
gauged WZW action.
Furthermore, the coset theory at level $k$ is known to describe
${\bf Z}_{k}$ parafermions and the corresponding deformation of this
conformal field theory by the first
thermal operator has been shown be integrable. It
has been conjectured that the resulting off-critical theory is
described by the minimal $a_{k-1}$ S-matrix \cite{fat}.
In this paper we will focus primarily on the semiclassical
quantization the CSG theory itself and will discuss these
connections to the coset model in detail elsewhere \cite{wip}.
\paragraph{}
The soliton solutions of the CSG equation carry the
charge, $Q$, associated with the global U$(1)$ symmetry of (\ref{lag}). After
quantization, $Q$ is restricted to integer values and the resulting
spectrum is a finite tower of massive charged particles.
Unlike the SG kink, the CSG soliton does not carry a topological
quantum number and there is no topological distinction between the vacuum
sector of the theory and the one-soliton sector. A surprising consequence of
this is that the elementary particle of the quantum CSG
theory can be identified with the $Q=1$ soliton itself.
In this paper we also calculate the semiclassical S-matrix
for the scattering of this particle with
with a soliton of arbitrary charge.
In particular we calculate this scattering amplitude
using both representations of the $Q=1$ particle of the theory
and find agreement, thus confirming the above mentioned
identification. As well as the leading-order behaviour of this
S-matrix element we are also able to calculate the leading-order
positions and residues of its poles.
These calculations provide us with several independent
semiclassical checks on our conjectured
exact S-matrix. The paper is organized as follows: in Section 2 we introduce
the CSG model and discuss the soliton solutions of the
field equation. We quantize these solutions and obtain the
semi-classical spectrum of the model. This section also contains a
calculation of the semiclassical phase shift in soliton-soliton scattering.
Most of the results in Section 2 are not new and were first obtained
in the paper by de Vega and Maillet \cite{dev2}, but we emphasize
several novel features of the model which have not been discussed
before. Section 3 is devoted
to a semiclassical calculation of the meson-soliton S-matrix.
In Section 4 we introduce our conjecture for the exact S-matrix of the
whole theory and show that the conjectured form
agrees with all the known semiclassical results for the theory
obtained in Sections 2 and 3. Finally we present our conclusions and
give a brief account of
the consistency of our results with the definition of the theory as a
deformed coset model mentioned above.
\section{Complex sine-Gordon solitons}
\paragraph{}
The classical equation of motion which follows from (\ref{lag}) is
\begin{equation}
\psi+\lambda^{2}\psi^{*}\frac{(\partial_{\mu}\psi)^{2}}{1-\lambda^{2}
|\psi|^{2}}+m^{2}\psi(1-\lambda^{2}|\psi|^{2})=0.
\label{ceq}
\end{equation}
This equation has a two parameter
family of time-independent soliton solutions:
\begin{equation}
\psi_{S}(x;\theta,X)=\frac{e^{i\theta}}{\lambda}\frac{1}{\cosh(m(x-X))}.
\label{tind}
\end{equation}
The symmetry coordinates $X$ and $\theta$ arise because the soliton solution
breaks the translational invariance and the global U$(1)$ invariance of the
Lagrangian (\ref{lag}). They correspond to the centre of mass of the soliton
and its orientation in the target space.
In addition, equation (\ref{ceq}) has explicit
time-dependent solutions,
labelled by a real parameter $\alpha$,
which correspond to a soliton rotating in
the internal U$(1)$ space with a constant angular velocity $\omega=
m\sin\alpha$; setting $\theta=X=0$ we have
\begin{equation}
\psi_{S}(x,t)=\frac{\cos(\alpha)}{\lambda}\frac{\exp(im\sin(\alpha)t)}
{\cosh(m\cos(\alpha)x)}.
\label{tdep}
\end{equation}
The static solution (\ref{tind}) is recovered by setting $\alpha=0$ in the
above formula. The classical mass $M$ and U$(1)$ charge, $Q$, are given by
\begin{eqnarray}
M=\int\, dx \frac{|\dot{\psi}|^2+|\psi'|^2}{1-\lambda^2|\psi|^2}
+m^{2}|\psi|^{2}, & \qquad{}
\qquad{} & Q=i\int\, dx\frac{\psi^{*}\dot{\psi}-\psi\dot{\psi}^{*}}
{1-\lambda^2|\psi^{2}|}.
\label{defs}
\end{eqnarray}
Evaluated on the time-dependent soliton solutions (\ref{tdep}) this yields
\begin{eqnarray}
M=\frac{4m}{\lambda^{2}}\cos(\alpha), & \qquad{} \qquad{} &
Q=\frac{4}{\lambda^{2}}\left({\rm sign}[\alpha]\frac{\pi}{2}-\alpha\right).
\label{mq}
\end{eqnarray}
\paragraph{}
It is instructive to note several unusual features of these solutions. Firstly
we note that both the mass and the charge of the soliton {\it decrease}
with the angular velocity of rotation $\omega$:
\begin{eqnarray}
M(\omega)=M(0)\sqrt{1-\frac{\omega^2}{m^{2}}}, & \qquad{} \qquad{}
& Q=\frac{4}{\lambda^{2}}\cos^{-1}(\omega),
\label{omega}
\end{eqnarray}
where $M(0)=4m/\lambda^{2}$ is the mass of the static soliton solution
(\ref{tind}).
This is in sharp contrast
with the more familiar behaviour of the Skyrmion, for which
\begin{eqnarray}
M(\omega)=M(0)+\frac{1}{2}\Lambda\omega^{2}, & \qquad{} \qquad{} &
Q=\Lambda\omega,
\label{skyrme}
\end{eqnarray}
where $\Lambda$ is the moment of inertia of the Skyrmion which is
independent of the angular velocity at weak coupling. In addition the moment
of inertia of the Skyrmion becomes large in the semi-classical limit and
correspondingly the rotational contribution to the soliton mass,
which goes like $Q^2/2\Lambda$, is small in this limit and always positive.
In the CSG case, the rotational contribution is
of the same order as the static contribution and is negative. In the extreme
case where the angular velocity reaches its maximum at $\alpha=\pi/2$ the
rotational contribution exactly cancels the static contribution and the mass
(as well as the soliton configuration itself) vanishes.
\paragraph{}
The second unusual feature
is the singular nature of the time-independent
solution obtained by setting $\alpha=0$.
The static solution (\ref{tind}) is the only solution which attains
the maximum value, $1/\lambda$; in this case at $x=X$.
At this value the factor
$1/(1-\lambda^2|\psi|^{2})$,
which appears the expressions (\ref{defs}) for the mass and
charge of the soliton, becomes singular. Although the mass of this field
configuration remains finite and equals $4m/\lambda^2$, its U$(1)$
charge is ill-defined. If we calculate the
charge $Q$ as function of $\alpha$ by
substituting (\ref{tdep}) for $\psi$ in (\ref{defs}), we find that this
function exhibits a branch cut which runs through the origin of the complex
$\alpha$-plane. For real $\alpha$ the result is given in equation (\ref{mq})
and we see that there is no unique limit as $\alpha\rightarrow 0$:
\begin{equation}
\lim_{\alpha\rightarrow 0^\pm}Q(\alpha)=\pm\frac{ 2\pi}{\lambda^{2}}.
\label{amb}
\end{equation}
This ambiguity suggests that the $U(1)$ charge
can only be defined in CSG theory modulo $4\pi/\lambda^{2}$.
This feature will play an important role in our proposed solution of the
model.
\paragraph{}
As for any Lorentz invariant field equation having soliton
solutions, for each of solutions given above we can generate a further
one-parameter family of uniformly moving
solutions by applying a Lorentz boost with rapidity
$\beta$. Hence the complete set of one soliton solutions of (\ref{ceq}) is
\begin{equation}
\psi_{S}(x,t;\alpha,\beta,\theta,X)=\frac{\cos(\alpha)}{\lambda}\frac{
\exp\left[i\theta+im\sin(\alpha)(\cosh(\beta)x-\sinh(\beta)t)\right]}
{\cosh\left[m\cos(\alpha)(\cosh(\beta)(x-X)-\sinh(\beta)t)\right]}.
\label{total}
\end{equation}
\paragraph{}
Having reviewed the classical properties of CSG solitons
we will now consider their quantization.
As we have explicit time-dependent soliton solutions we can obtain the
semiclassical spectrum of soliton states by applying the Bohr-Sommerfeld
quantization rule; $S[\psi_{S}]+M\tau=2\pi n$, where $n$ is a positive
integer, $\tau=2\pi/\omega$ is the
period of the  soliton solution (\ref{tdep}) and $S$ is its action; whence,
\begin{equation}
S[\psi_{S}]+M\tau=\int_{0}^{\tau}\,dt\int_{-\infty}^{+\infty}\,dx
\frac{2|\dot{\psi_{S}}|^{2}}{1-\lambda^{2}|\psi_{S}|^{2}}=2\pi|Q|,
\end{equation}
and therefore the charge $Q$ is restricted to integer values.
Hence the semiclassical mass spectrum
of the charged soliton states is given by,
\begin{equation}
M(Q)=\frac{4m}{\lambda^2}\left|\,
\sin\left(\frac{\lambda^{2}Q}{4}\right)\right|,
\label{spectrum}
\end{equation}
where $Q=\pm 1,\pm 2,\ldots,\pm Q_{\rm max}=
[2\pi/\lambda^2]$ and $[x]$ denotes
the greatest integer less than or equal to $x$. The above spectrum is a
periodic
function of $Q$ with period $4\pi/\lambda^2$ which again suggests
that it is correct
to think of $Q$ as an angular variable.
The spectrum is similar in form to the
spectrum of breather states in the ordinary sine-Gordon theory except here
the states are labelled by a conserved U$(1)$ charge which takes negative
as well as  positive values. The lightest states in the spectrum have charge
$Q=\pm 1$ and mass $M(1)=m+O(\lambda^4)$. These quantum numbers are identical
to those of the elementary meson of the theory
(together with its anti-particle)
which is interpolated by the complex scalar field $\psi$
and we therefore assume that the $Q=\pm 1$ state {\it is} the
elementary meson. The state of charge $Q$ then has an obvious interpretation
as a bound-state of $Q$ mesons.
A similar phenomenon occurs for the the breathers of the SG
theory, where the lightest breather state is identified with
the elementary particle of that theory. Thus, in that case, the elementary
particle is identified as a bound state of two sine-Gordon kinks.
In the CSG case,
we have just found that the elementary particle is to be identified as a
rotational excitation of the soliton itself. This is possible because the
CSG soliton does not carry a topological charge and correspondingly
there is no topological distinction between the vacuum sector and the
one-soliton sector.
\paragraph{}
In general each state of charge $Q$ has a partner of charge $-Q$ which is
identified as its anti-particle. However, an important special case
occurs when the coupling constant takes one of the the infinite
discrete set of special
values $\lambda^{2}=2\pi/N$ where $N=1,2,\ldots$.
At this point the number of stable particles
in the spectrum changes discontinuously from $2N-2$ to $2N$. The
Bohr-Sommerfeld quantization rule means that each state in the quantum theory
is associated with a single time-dependent solution of the CSG field
equation which has the same mass and charge. However, precisely at these
special values the two soliton states with charge $Q=\pm Q_{\rm max}=\pm N$
are associated with the {\em same} solution; the static soliton (\ref{tind}).
As discussed above the U$(1)$ charge of this classical
solution is double-valued; having an ambiguity of exactly $2N$. We
propose a resolution fo this puzzle; a single valued spectrum is
obtained by identifying U(1) charges modulo $2N$ so that the states
$\pm N$ actually correspond to the same particle. More generally for
the values $\lambda^{2}=4\pi/k$ where $k$ is an integer $>1$, the U(1)
charge is identified modulo $k$ and therefore should more properly be
thought of as a ${\bf Z}_k$ charge. In the following we will show that
this assumption leads to a completely consistent quantum theory.
\paragraph{}
One-loop quantum corrections to the semiclassical spectrum (\ref{spectrum})
were computed by de Vega and Maillet \cite{dev2}. They found that the only
effect of these corrections was an overall finite renormalization of the
coupling constant:
\begin{eqnarray}
\lambda^{2} & \rightarrow &
\lambda^{2}_{R}=\frac{\lambda^{2}}{1-\lambda^{2}/4\pi}.
\label{renorm}
\end{eqnarray}
This finite renormalization of the coupling is familiar from the
SG theory where the resulting spectrum turns
out to be exact. In \cite{dev2} it is conjectured
that, provided suitable finite counter-terms are added to the action to
maintain integrability at the quantum level, the one-loop spectrum for the CSG
theory is also exact. In the following, we will assume that this is the case
and we will also assume that the
S-matrix describing the scattering of the charged solitons described above
is exactly factorizable. We will show that these assumptions imply a certain
minimal form for the S-matrix of the theory and check that this minimal form
passes a series of stringent semiclassical tests. However, we shall
only be able to find a consistent S-matrix for which the bootstrap
closes when the renormalized coupling constant takes one of the
special values described above; $\lambda^{2}_{R}=4\pi/k$.
\paragraph{}
The classical CSG equation is completely integrable and explicit solutions
which describe the scattering of two solitons of arbitrary charge can be
constructed, either by the method of Hirota \cite{get}, or by the inverse
scattering method \cite{dev2}. As mentioned above, the solitons undergo
completely elastic scattering with no change in shape, the only result of
the collision being a time delay.
The analytic expressions for these solutions are extremely cumbersome, we
refer the reader to \cite{dev2} for details. However, the time delay for
soliton scattering can be extracted easily from the large-time asymptotics of
these expressions. Consider two solitons with rapidities $\beta_{1}$,
$\beta_{2}$ and internal parameters $\alpha_{1}$ and $\alpha_{2}$. The
centre-of-momentum (COM)
frame for the two solitons can be chosen by imposing the
relation; $\sinh(\beta_{1})\cos(\alpha_{1})=-\sinh(\beta_{2})
\cos(\alpha_{2})$. In the COM frame both solitons experience the same time
delay due to the collision:
\begin{equation}
\Delta t(\beta_{1},\beta_{2},\alpha_{1},\alpha_{2})
=\frac{2}{m\sinh(\beta_{1})\cos(\alpha_{1})}\log\left |\frac
{\sinh\left(\Delta\beta+i\Delta\alpha\right)}{\cosh
\left(\Delta\beta+i\bar{\alpha}\right)}\right |,
\label{delay}
\end{equation}
where $\Delta\beta=(\beta_{1}-\beta_{2})/2$, $\Delta\alpha=(\alpha_{1}-
\alpha_{2})/2$ and $\bar{\alpha}=(\alpha_{1}+\alpha_{2})/2$. The
semi-classical phase shifts $\delta(E)$
for soliton-soliton scattering at total COM energy E, are completely
determined by the time-delay via the WKB formula:
\begin{equation}
\delta(E)=\frac{1}{2}n_{B}\pi+
\frac{1}{2}\int_{E_{\rm Th}}^{E} dE'\, \Delta t(E'),
\label{wkb}
\end{equation}
where $n_{B}$ is the number of quantized breather states
below the threshold energy, $E_{\rm Th}$.
As there are no known breather solutions of the CSG equation
we will assume that $n_{B}=0$ \footnote{Because
the equations of motion (\ref{ceq}) depend on $\psi$ and $\psi^\star$,
breather solutions cannot
be obtained by analytic continuation to imaginary values of
$\Delta\beta$ as is the case in ordinary SG theory. On this point we
are in disagreement with \cite{get}.}.
This semi-classical formula for the phase shift was derived in the context of
soliton scattering by Jackiw and Woo \cite{jac}. The COM energy of the
two soliton scattering solution is just given by the sum of the
two soliton energies which is conveniently written as
\begin{equation}
E=\frac{4m}{\lambda^2}(\cos(\alpha_{1})\cosh(\bar{\beta}+\Delta\beta)+
\cos(\alpha_{2})\cosh(\bar{\beta}-\Delta\beta)),
\label{com}
\end{equation}
with $\bar{\beta}=(\beta_{1}+\beta_{2})/2$. Hence the semiclassical
phase shift is given by;
\begin{equation}
\delta(\Delta\beta)=\frac{4}{\lambda^{2}}
\int_{0}^{\Delta\beta}dx\,\log\left[\frac{\sinh(x+i\Delta\alpha)
\sinh(x-i\Delta\alpha)}{\cosh(x+i\bar{\alpha})\cosh(x-i\bar{\alpha})}\right].
\label{I1}
\end{equation}
It is easily checked that this expression agrees exactly with formula (4.8) of
ref \cite{dev2} which was derived by slightly different means. An
alternative and, for our purposes, more convenient
integral representation for the semiclassical phase shift
can be derived by changing the contour of integration in
(\ref{I1}):
\begin{equation}
\delta(\Delta\beta)=-\frac{2i}{\lambda^{2}}
\int_{2\Delta\alpha}^{\pi-2\bar{\alpha}}d\eta\,
\log\left[\frac{e^{i\eta+2\Delta\beta}-1}{e^{i\eta}-e^{2\Delta\beta}}\right].
\label{I2}
\end{equation}
\section{Meson-soliton scattering}
\paragraph{}
The aim of our investigation is to calculate the S-matrix for the charged
particles described in the previous section. Since the
$Q=1$ particles can be represented either as solitons or as
elementary mesons means that we can apply two separate methods for
calculating the two-body scattering amplitude for these particles with the
states of higher charge. Firstly, the
result (\ref{I1}) of the previous section and the formula (\ref{mq})
for the soliton charges imply that the semiclassical
phase shift for meson soliton scattering is given by
\begin{equation}
\delta(\theta)=-\pi-\frac{2i}{\lambda^{2}}
\int_{\frac{\lambda^{2}}{4}(Q-1)}^{\frac{\lambda^{2}}{4}(Q+1)}
d\eta\,\log\left[\frac{\sinh\left(\frac{\theta}{2}+i\frac{\eta}{2}\right)}{
\sinh\left(\frac{\theta}{2}-i\frac{\eta}{2}\right)}\right].
\label{phase1}
\end{equation}
where $\theta=\beta_{1}-\beta_{2}$. The usual
simplifying feature of the semi-classical limit for meson-soliton scattering
is that the soliton becomes very massive. However in the CSG case this is not
true for all the states in the semi-classical spectrum (\ref{spectrum}).
For this reason
we will restrict our attention to the case where the mass of the charge-$Q$
particle grows like $\lambda^{-2}$ in the semi-classical limit. This means
choosing a value of $Q$ which is remains a
finite distance from the top of the tower
(\ref{spectrum}) in this limit; hence we take $Q\sim\lambda^{-2}$. The
leading semi-classical behaviour of the S-matrix,
${\cal S}(\theta)=\exp(2i\delta)$ is then given by the simple formula
\begin{equation}
{\cal S}_{1,Q}(\theta)=\frac{\sinh^{2}\left(\frac{\theta}{2}+
i\frac{\lambda^2 Q}{8}\right)}{\sinh^{2}\left(\frac{\theta}{2}-
i\frac{\lambda^2 Q}{8}\right)}+O(\lambda^2).
\label{smatrix1}
\end{equation}
\paragraph{}
In fact it is also possible to obtain this formula for ${\cal S}(\theta)$
directly from the mesonic representation of the $Q=1$ particle. Let us expand
the field around the rest-frame soliton solution $(\ref{tdep})$ as
$\psi=\psi_{S}(x;\alpha)+\delta\psi$. The Lagrangian density becomes
\begin{equation}
{\cal L}[\psi]={\cal L}[\psi_{S}]+\delta\psi^{*}\hat{M}\delta\psi+
O(\delta\psi^{3}),
\label{fluctuation}
\end{equation}
where
$\hat{M}=(\delta^{2}{\cal L}/\delta\psi\delta\psi^{*})[\psi_{S}(x;\alpha)]$
is a Hermitian differential operator. As is well
known, the continuous part of the spectrum of $\hat{M}$ corresponds to the
scattering of a single meson off the background soliton field configuration.
The phase shifts of these scattering eigenmodes have been found by de Vega
and Maillet using the inverse scattering method. In our notation, their
equation (3.18a) for the phase shift of a positively charged meson
scattering off a soliton of charge $Q$ reads
\begin{equation}
\delta_{+}(\theta)=2\tan^{-1}
\left[\frac{\sin\left(\frac{\lambda^{2}Q}{4}\right)
\sinh(\theta)}{\cos\left(\frac{\lambda^{2}Q}{4}\right)\cosh(\theta)-1}\right].
\label{dv1}
\end{equation}
It is straightforward to check that the
resulting S-matrix element, $\exp(2i\delta_{+})$, coincides exactly with
the formula (\ref{smatrix1}) obtained above. This
agreement provides further confirmation for the identification of the $Q=1$
soliton state with the elementary meson of the theory.
\paragraph{}
Although the leading order $S$-matrix (\ref{smatrix1}) for meson-soliton
scattering is unitary and completely elastic by construction it lacks several
features we would expect from the exact S-matrix of the theory. In
particular, given the existence of a stable particles of charge $Q+1$ and
$Q-1$ in the semiclassical spectrum (\ref{spectrum}) for $0<Q<Q_{\rm max}$,
we would expect these particles to arise as intermediate states
in the scattering
amplitude in the direct and crossed channel respectively. Assuming the
one-loop semiclassical spectrum is exact,
the kinematics of the processes illustrated
in Figs 1 a) and b)
dictates the presence of simple poles in the scattering amplitude
${\cal S}_{1,Q}(\theta)$ at $\theta=i\lambda_{R}^{2}(Q\pm 1)/4$ (recall that
$\lambda^2_{R}=\lambda^{2}+O(\lambda^{4})$).
In fact, rather
than these simple poles, the leading semi-classical expression has a double
pole at $\theta=i\lambda^{2}Q/4$. The reason for this is simply that for a
soliton state a fixed distance from the top of the tower with charge
$Q\sim\lambda^{-2}$ the splitting between the direct and crossed channel poles
is a sub-leading effect; at leading order these poles merge to form a double
pole.
\paragraph{}
If we assume that the exact S-matrix for meson-soliton scattering is
factorizable then its form is highly constrained (for a review of
elastic S-matrices in two dimensions see \cite{zam}).
In particular, the absence of particle creation and of reflection
for particles of different masses means that the unitarity condition
takes the simple form: ${\cal S}(\theta){\cal S}(-\theta)=1$. This
condition together with the assumption of
analyticity implies that the S-matrix must be a
product of factors of the form
\begin{equation}
F_{x}(\theta)=\frac{\sinh\left(\frac{\theta}{2}+i\frac{\lambda_{R}^{2}x}{8}
\right)}
{\sinh\left(\frac{\theta}{2}-i\frac{\lambda_{R}^{2}x}{8}\right)},
\label{block}
\end{equation}
where we have chosen a convenient normalization for the label $x$.
Each factor has a simple pole at $\theta=ix\lambda_{R}^{2}/4$.
Clearly the minimal
choice for the S-matrix which has the two required simple poles is
\begin{equation}
{\cal S}_{1,Q}(\theta)=F_{Q-1}(\theta)F_{Q+1}(\theta).
\label{conjecture}
\end{equation}
It is easy to check that this expression agrees with our semiclassical
result (\ref{smatrix1}) at leading order in $\lambda^2$. In the following we
will provide further evidence that (\ref{conjecture}) is the exact S-matrix
for meson soliton scattering in CSG theory.
\paragraph{}
As discussed above the simple poles in the meson-soliton scattering amplitude
correspond to the process where the soliton undergoes a transition from its
excited state of charge $Q$ to the state of charge $Q\pm 1$ by absorbing or
emitting a single meson. These processes are completely analogous to the
contribution of the $\Delta$ resonance to pion-nucleon scattering in the
Skyrme model \cite{adk,dor1}.
Of course, in the present case, the mass splitting between adjacent
soliton states in the tower is always less than the meson mass, so these
processes are below threshold and can only be realized at imaginary values
of the external momenta. In the Skyrme model,
although the mass splitting between the
$\Delta$ and the nucleon is a subleading effect in the semi-classical limit,
the pion mass is an independent parameter of the theory, and hence the
$\Delta$
resonance can be kept above threshold. Because the splitting is subleading,
a straightforward semi-classical analysis of the pion-Skyrmion S-matrix
by finding the continuous spectrum of the small fluctuation operator, exactly
analogous to the derivation of the result (\ref{dv1}),
does not find the $\Delta$ contribution. In recent work, one of the
authors \cite{dor1}
has developed a method which allows a systematic semi-classical evaluation
of the width of the $\Delta$ resonance. The method is equally applicable to
processes below threshold and can be used to calculate
the positions and residues of the S-matrix poles corresponding to the
transitions $Q\rightarrow Q\pm 1$ providing a further check on the proposed
exact S-matrix. The first step is to calculate the leading semiclassical
contribution to the two-point Green's function of the meson field in the
one-soliton sector. This can be accomplished by applying the saddle-point
approximation to an appropriate Feynman path integral; some details of this
calculation are given in the Appendix.
The bound-state contribution  to the T-matrix can then be obtained by applying
the LSZ reduction formula to this Green's function, the result is
\begin{equation}
{\cal T}_{1,Q}(\omega)=-\frac{16m^2}{\lambda^{2}}\left[\frac{\sin^{4}\left(
\frac{\lambda^{2}\left(Q+\frac{1}{2}\right)}{4}\right)}{\omega-M(Q+1)+M(Q)}
-\frac{\sin^{4}\left(\frac{\lambda^{2}\left(Q-\frac{1}{2}\right)}{4}\right)}
{\omega-M(Q)+M(Q-1)}\right],
\label{final}
\end{equation}
where $\omega$ is the energy of the incoming meson in the rest frame of the
soliton. The two terms in this expression correspond directly to the processes
shown in Figures 1 a) and 1 b) respectively.
With the correct normalization \cite{kaw}
the corresponding contribution to the S-matrix is given by
\begin{equation}
{\cal S}_{1,Q}=1+\frac{{\cal T}_{1,Q}(\omega)}{2ik},
\label{st}
\end{equation}
where $k$ is the meson momentum.
We now compare this result with our conjecture for the exact meson soliton
S-matrix
(\ref{conjecture}). As we are working in the rest frame of the soliton,
the rapidity difference $\theta$ is simply the rapidity of the
incoming meson; thus we set $\omega=m\cosh(\theta)$ and $k=m\sinh(\theta)$.
{}From (\ref{final}) we find that the S-matrix has poles at
$\theta=i\lambda^2(Q\pm \frac{1}{2})/4$ with residues
\begin{equation}
{\cal R}_{\pm}=\pm \frac{8i}{\lambda^2}\sin^{2}
\left(\frac{\lambda^2\left(Q\pm\frac{1}{2}\right)}{4}\right).
\label{residue}
\end{equation}
As mentioned above the correct positions of the poles are dictated by
relativistic kinematics to be at $\theta=i\lambda_{R}^{2}(Q\pm 1)/4$. The
discrepancy with the saddle-point result is due to our neglect of the
translational motion of the soliton and the renormalization of the
coupling, both of which are
subleading effects as long as
$Q\sim\lambda^{-2}$. The proposed exact result has a poles
at the above mentioned positions with residues ${\cal R}^{exact}_{\pm}=\pm
(8i/\lambda_{R}^{2})\sin^{2}(\lambda_{R}^{2}2 Q/4)$.
Clearly the pole positions and
residues agree to leading order in the semi-classical limit $\lambda^2
\rightarrow 0$.
\section{Soliton-soliton scattering}
\paragraph{}
In the previous section we found evidence to support a completely
factorizable form (\ref{conjecture})
for the scattering amplitude of a soliton of unit charge
with a soliton of charge $Q\sim \lambda^{-2}$.
This encourages us to extend this conjecture
to the scattering amplitude for two solitons of arbitrary charges.
As discussed in the previous section, unitarity and
analyticity imply that the factorizable S-matrix element must be a product of
factors $F_{x(i)}(\theta)$ defined in (\ref{block}), for some set of
real numbers $\{x(i)\}$. As each of these factors supplies a simple pole, the
minimal choice for the S-matrix is determined by the set of
singularities which are consistent with the spectrum (\ref{spectrum}) (with
the one-loop replacement $\lambda^{2}\rightarrow \lambda^{2}_{R}$).
\paragraph{}
Let us first consider the case of two solitons with charges
$Q_{1}\geq Q_{2}>0$ such that $Q_{1}+Q_{2}\leq Q_{\rm max}$. In this case the
semi-classical spectrum tells us that there are stable particles with
charges $Q_{1}+Q_{2}$ and $Q_{1}-Q_{2}$ and we expect these states to
arise as intermediate states in the direct and crossed channels respectively
(see Figs 2 a) and b)). It follows that there should be simple poles in the
scattering amplitude at $\theta=i\lambda_{R}^{2}(Q_{1}\pm Q_{2})/4$
and thus we
include the factors $F_{Q_{1}+Q_{2}}(\theta)$ and $F_{Q_{1}-Q_{2}}(\theta)$.
However, simple poles corresponding to bound states are not the only
kinematically allowed singularities of the S-matrix. In four dimensions,
anomalous thresholds due to two-particle intermediate states lead to cuts in
the the amplitude, in two dimensions these singularities become double poles.
The rules for determining the positions of these double poles were discussed
in detail by Coleman and Thun \cite{col}. In particular, these authors used
their rules to explain the positions of double poles in the
breather S-matrix of the SG model. In the CSG case the
kinematics of the processes illustrated in Fig 4, in which the intermediate
state consists of two solitons with charges $Q_{2}-n$ and
$Q_{1}+n$ for $n=1$,$2$,$\ldots$,$Q_{2}-1$, leads to a total of
$Q_{2}-2$ double poles in the amplitude at
$\theta=i\lambda^{2}_{R}(Q_{1}-Q_{2}+2n)/4$. The minimal choice for a
factorizable S-matrix which has all the required poles is therefore given by
\begin{equation}
{\cal S}_{Q_{1},Q_{2}}(\theta)=F_{Q_{1}-Q_{2}}(\theta)\left[
\prod^{Q_{2}-1}_{n=1}
F_{Q_{1}-Q_{2}+2n}(\theta)\right]^{2}F_{Q_{1}+Q_{2}}(\theta).
\label{smatrix2}
\end{equation}
\paragraph{}
The particles in the semi-classical spectrum occur in pairs of opposite
charge and it is natural to interpret the state with charge $-Q$ as the
anti-particle of the state with charge $Q$. In this case the extension of our
proposed S-matrix to negative values of $Q_{1}$ and $Q_{2}$ is completely
determined by crossing symmetry:
\begin{eqnarray}
{\cal S}_{Q_{1},-Q_{2}}(\theta) & = &
{\cal S}_{Q_{1},Q_{2}}(i\pi-\theta), \nonumber \\
{\cal S}_{-Q_{1},-Q_{2}}(\theta) & = &
{\cal S}_{Q_{1},Q_{2}}(\theta).
\label{crossing}
\end{eqnarray}
\paragraph{}
We will now check that the above S-matrix reduces to the semi-classical
expression derived in Section 1 above. Unlike the previous section, we will
consider two solitons of fixed U(1) charge in the semiclassical
limit: $Q_{1}$, $Q_{2}$ $\sim\lambda^{0}$.
We first consider the case $Q_{1}
\geq Q_{2}>0$. Using the integral
representation (\ref{I2}) for the semiclassical phase shift we find
\begin{equation}
{\cal S}_{Q_{1},Q_{2}}(\theta)=
\exp\left(\frac{4}{\lambda^2}\int_{\frac{\lambda^2}{4}(Q_{1}-Q_{2})}
^{\frac{\lambda^{2}}{4}(Q_{1}+Q_{2})}d\eta \log\left[
\frac{\sinh\left(\frac{\theta}{2}+i\frac{\eta}{2}\right)}
{\sinh\left(\frac{\theta}{2}-i\frac{\eta}{2}\right)}\right] \right)\,+
\,O(\lambda^{4}).
\label{semic}
\end{equation}
Defining the discrete set of points
$\eta_{j}=\lambda^{2}(Q_{1}-Q_{2})/4+j\lambda^{2}/2$, for integers $j$
with $0\leq j\leq Q_{2}$,
the definition of the Riemann integral of an arbitrary function $f(\eta)$ is
\begin{equation}
\int_{\frac{\lambda^{2}}{4}(Q_{1}-Q_{2})}^{\frac{\lambda^2}{4}(Q_{1}+Q_{2})}
d\eta\, f(\eta)=\frac{\lambda^2}{4}
\sum_{j=1}^{Q_{2}}\left[f(\eta_{j})+f(\eta_{j-1})\right]\,+\,O(\lambda^{6}).
\label{riemann}
\end{equation}
The equality of our proposed S-matrix $\cal S$ and the semiclassical
result (\ref{semic}) to $O(\lambda^4)$ follows immediately from setting
$f(\eta)=\log(F_{\frac{4\eta}{\lambda^{2}}}(\theta))$
in the above relation. This agreement
can immediately be extended to the cases $Q_{1}<0$ and/or $Q_{2}<0$ by
noting that the semiclassical phase shift (\ref{I2}) satisfies the
crossing relationship,
\begin{equation}
\delta_{Q_{1},-Q_{2}}(\theta)=\delta_{Q_{1},Q_{2}}(i\pi-\theta)
\label{crossing2}
\end{equation}
\paragraph{}
To obtain the complete S-matrix of the theory it is necessary to relax our
requirement that $Q_{1}+Q_{2}\leq Q_{\rm max}$. In fact the agreement between
our conjectured S-matrix and the semiclassical expression (\ref{semic})
extends to all values of $Q_{1}$ and $Q_{2}$ and, in this sense, there is
no obstacle to applying the formulae (\ref{smatrix2}) everywhere.
However, there is a now a problem in the sense that the S-matrix
elements ${\cal S}_{Q_1,Q_2}(\theta)$ with $Q_1+Q_2>Q_{\rm max}$ have
a simple pole at $\theta=i\lambda_{R}^{2}(Q_{1}+Q_{2})/4$ which cannot
be explained by the exchange in either the direct or crossed channels
of one the solitons in the spectrum. Either the spectrum must be
enlarged to include new states or these simple poles must be accounted
for via another mechanism. We will defer a discussion of this problem
till the last section.
\paragraph{}
Fortunately, in the special cases discussed in
Section 1, where $\lambda_{R}^{2}=4\pi/k$ for some integer $k>1$,
we will see that all the
poles in the S-matrix (\ref{smatrix2}), for arbitrary values of
$Q_{1}$ and $Q_{2}$,
{\em can} be explained in terms of the particles
in the semiclassical spectrum alone. This explanation implies that the
U(1) charge is only conserved modulo $k$ which is directly related to
the ambiguity (\ref{amb}). In this case the spectrum consists of $k-1$
particles with $Q=\pm1,\pm2,\ldots,\pm(k-1)/2$ for $k$ odd and
$Q=\pm1,\pm2,\ldots,\pm(k-2)/2,k/2$ for $k$ even. (In the later case
recall that the states with charge $\pm k/2$ are identified.)
The masses of these particles are given by
\begin{equation}
M(Q)=\frac{km}{\pi}\left|\sin\left(\frac{Q\pi}{k}\right)\right|.
\label{toda}
\end{equation}
This is the spectrum of the minimal S-matrix
associated with the Lie algebra $a_{k-1}$ \cite{kob} and indeed, for
$\lambda^{2}=4\pi/k$, our conjecture (\ref{smatrix2}) is precisely
that S-matrix.
In this case, for any $Q_{1}$ and $Q_{2}$,
there is a always two particles in the spectrum (\ref{toda}) with charge
$Q_{3}=Q_{1}\pm Q_{2}$ mod$\,k$ which correspond to the simple poles in
$S_{Q_{1},Q_{2}}(\theta)$. In fact, it is well known that the
minimal $a_{k-1}$ S-matrix defines a completely consistent theory
in which all poles, both simple and double, can be explained in terms of the
above spectrum (see for example \cite{bra}).
\section{Conclusions}
\paragraph{}
The main result of this paper is as follows: if one assumes that the CSG
Lagrangian with $\lambda_{R}^{2}=4\pi/k$
defines a completely integrable quantum field theory for every
integer $k>1$, then
the minimal $a_{k-1}$ S-matrix is the minimal S-matrix
which is consistent with the semiclassical spectrum of the model.
We therefore conjecture that this S-matrix together with the one-loop
semiclassical spectrum constitute the exact solution of the CSG model for
these values of the coupling.
Our conjectured S-matrix successfully reproduces the leading
semiclassical
behaviour of the scattering of two solitons of arbitrary charges. In addition
the conjectured form agrees both with the
leading semiclassical expression for the
meson-soliton scattering amplitude and also with the leading order
calculation of the positions and residues of the poles in this
amplitude.
These semiclassical checks clearly imply that any non-minimal factors in the
exact S-matrix must vanish at $O(\lambda^{2})$. It is then straightforward to
argue that any CDD factors which are consistent with the $a_{k-1}$
bootstrap equations cannot be present. For instance,
for the scattering of two particles of charge $Q=1$, the CDD factor
which is consistent with the bootstrap equations for $a_{k-1}$ has the
form
\begin{equation}
{\sinh\left(\frac{\theta}{2}-\frac{i\pi}{2k}(2-\alpha)\right)\sinh\left(
\frac{\theta}{2}-\frac{i\pi}{2k}\alpha\right)
\over
\sinh\left(\frac{\theta}{2}+\frac{i\pi}{2k}(2-\alpha)\right)
\sinh\left(\frac{\theta}{2}+\frac{i\pi}{2k}\alpha\right)},
\end{equation}
where $\alpha(\lambda^2)$ is some function of the coupling constant
such that $0{<}\alpha{<}2$, in order that no additional poles are
introduced onto the physical strip.
Notice that such a CDD factor always has a contribution at $O(1/k)$,
i.e. $O(\lambda^2)$, whatever the functional dependence of $\alpha$,
and is therefore ruled out by our tree-level results.
\paragraph{}
A final possible semiclassical check on our proposed S-matrix is to calculate
the scattering amplitude for two mesons using ordinary
perturbation theory and compare it with our conjectured form
${\cal S}_{1,1}(\theta)=F_{2}(\theta)$. This scattering
amplitude has been calculated to one loop by de Vega and Maillet
\cite{dev1}. Our conjecture agrees with their calculation at
$O(\lambda^{2})$ but differs at $O(\lambda^{4})$. However, the one-loop
contribution of Ref \cite{dev1} depends crucially on the form of the
finite counter-terms added to the action and
we believe that these terms should be re-evaluated in the light
of the definition of the model as a perturbed coset theory.
\paragraph{}
For other values of the coupling the S-matrix elements have simple
poles which would seem to require at least
the existence of additional states in order to be explained. These
additional states are not realized semiclassically; for instance, we have
argued that there are no breather solutions. Another
possibility is that the scattering of a soliton $Q$ with $-Q$ has a reflection
amplitude away from the discrete values of the coupling. This idea is
motivated by analogy with the SG theory where the soliton S-matrix in general
is non-diagonal, except at the threshold values of the coupling where
the total S-matrix for the solitons and breathers
becomes the minimal $d_k$ S-matrix, up to signs \cite{bra}. However,
after pursuing these two possibilities, we have not found it possible
to define a consistent factorizable
S-matrix for these other values of the coupling constant.
As we mentioned in the introduction this has a perfectly natural
explanation in terms of the realization of the model as a
perturbed coset model which is only
well-defined quantum mechanically if the level $k$ is an integer $>1$.
The perturbation of the SU$(2)/{\rm U}(1)$ coset model at level $k$ by
the first thermal operator having dimensions $\Delta=\bar\Delta=2/(k+2)$
has been discussed in \cite{fat}. Following the methods of
Zamolodchikov \cite{zamint} the perturbation is shown to be integrable and
the S-matrix has been conjectured to be the $a_{k-1}$ minimal
S-matrix. This is precisely the result of our calculation. Notice that
the hidden ${\bf Z}_k$ structure of the solitons manifests the ${\bf
Z}_k$ structure of the parafermions, a fact which deserves to be
studied in more detail. Finally we note that the part of the finite
renormalization of the coupling (\ref{renorm}) which comes from purely
from the Lagrangian (\ref{lag}) {\footnote{In other words, omitting the
contribution of the finite one-loop counter-term \cite{dev2}.}}
corresponds exactly to an integer shift in the level $k\rightarrow
k+2$ which is encountered in gauged
WZW models (see \cite{tsey} and references therein).
A detailed discussion
of this correspondence will be presented elsewhere \cite{wip}.
\paragraph{}
TJH would like to thank Ioannis Bakas for first bringing the CSG theory to
his attention and for many discussions, and Q-Han Park for some useful
conversations. The authors would like to thank Marco Kneipp for a useful
contribution to one of the calculations in this paper. ND thanks the
Theory Division at CERN for hospitality while part of this work
was being completed. TJH would like to acknowledge the support of CERN
where this work was begun.
\section*{Appendix}
\paragraph{}
The purpose of this Appendix is to calculate the positions and
residues of the simple poles in the meson-soliton S-matrix to leading
order in the semiclassical approximation. The calculation is
essentially a direct application of the method developed in
\cite{dor1} and we refer the reader to this paper for further details.
The basic object in quantum field theory from which the meson-soliton
scattering amplitude can be extracted is a two-point Green's function for the
meson field
$\psi$ evaluated not in the vacuum sector but rather in the one-soliton state
of charge $Q$:
\begin{equation}
G_{Q}(x,x';t-t')
=\langle Q|\psi(x,t)\psi^{*}(x',t')|Q \rangle.
\label{defg}
\end{equation}
Following the collective coordinate approach of Gervais, Jevicki
and Sakita \cite{ger}, we
write this Green' function as a phase-space path integral over
$\psi$ and its conjugate momentum;
$\pi=\delta {\cal L}/\delta \dot{\psi}^{*}$.
\begin{equation}
G_{Q}=\int\, {\cal D}\psi{\cal D}\psi^{*}{\cal D}\pi{\cal D}\pi^{*}
\Phi_{Q}^{*}[\psi(x,T)]\Phi_{Q}[\psi(x,-T)] \psi(x,t)\psi^{*}(x',t')
\exp\left(i\int\, d^{2}x \pi\dot{\psi}^{*}+\pi^{*}\dot{\psi}-{\cal H} \right),
\label{tfi}
\end{equation}
where $\Phi_{Q}$ is the wave-functional for the soliton state of charge $Q$
which will be identified explicitly below.
The Hamiltonian density is given by
${\cal H}=\pi^{*}g^{-1}(|\psi|)\pi+V(|\psi|,|\psi'|)$ where
\begin{eqnarray}
g(|\psi|)=\frac{1}{1-\lambda^{2}|\psi|^{2}} & \qquad{} \qquad{}
& V(|\psi|,|\psi'|)=g(|\psi|)|\psi'|^{2}+m^2|\psi|^{2}.
\end{eqnarray}
The CSG Lagrangian differs from the model considered in \cite{dor1}
through the presence of the factor $g(|\psi|)$ which appears in the
kinetic term. The generalization of the analysis to include exactly
such a factor, which can be thought of as a target-space metric, was
given in Appendix A of \cite{dor2}.
The essential result of reference \cite{dor1} is that the leading contribution
of the intermediate states $|Q\pm 1\rangle$ to meson-soliton scattering can
be extracted by evaluating this path-integral in the
semi-classical saddle-point approximation.
The saddle-point equation is
just the classical field equation itself and so the relevant saddle-point
field configurations $\psi_{sp}$
are the time-dependent one-soliton solutions (\ref{tdep}) which are
parametrized by a collective coordinate $\theta$ and its conjugate momentum
which is the $U(1)$ charge, $Q$:
\begin{equation}
\psi_{sp}(x,t;\theta,Q)=\frac{e^{i\theta}}{\lambda^2}\frac{\sin\left(\frac
{\lambda^{2}Q}{4}\right)}{\cosh\left(m\sin\left(\frac{\lambda^{2}Q}{4}\right)
x\right)}.
\label{sp}
\end{equation}
As in reference \cite{dor1}, we have neglected the translational motion of the
soliton which gives subleading recoil corrections to the S-matrix elements.
At leading order the wave-functional for the state $|Q\rangle$ is a
function of the collective coordinate $\theta$ only; up to an irrelevant
normalization $\Phi_{Q}=\exp(iQ\theta)$ and the saddle-point contribution to
the path integral is just
\begin{eqnarray}
G_{Q}&=& \int {\cal D}\theta(t) {\cal D}Q(t)\exp[iQ(\theta(-T)-\theta(T))]
\psi_{sp}(x,t;\theta(t),Q(t))\psi^{*}_{sp}(x',t';\theta(t'),Q(t'))
\nonumber \\
& & \qquad{} \qquad{} \qquad \qquad{} \qquad{} \qquad{} \times \,
\exp\left(\int_{-T}^{T}dt\,Q\dot{\theta}-\frac{4m}{\lambda^{2}}\sin\left(\frac
{\lambda^{2}Q}{4}\right)\right).
\label{qm}
\end{eqnarray}
Now the problem has been reduced to one-dimensional quantum mechanics of the
Hamiltonian $\hat{H}=(4m/\lambda^{2})\sin(\lambda^{2}\hat{Q}/4)$
with the canonical commutation relation $[\hat{\theta},\hat{Q}]=i$. Thus,
\begin{equation}
G_{Q}(x,x';t-t')  =  \langle Q|\psi_{sp}(x,t;\hat{\theta},\hat{Q})
\psi^{*}_{sp}(x',t';\hat{\theta},\hat{Q})|Q\rangle,
\label{operator}
\end{equation}
where the operator ordering ambiguity inherent in the saddle-point field
operator
$\hat{\psi}_{sp}(x,t)=
\psi_{sp}(x,t;\hat{\theta},\hat{Q})$ is resolved by choosing the Weyl
ordering prescription \cite{dor1}. It is easy to see that this operator,
being proportional to $\exp(i\hat{\theta})$,
only gives a non-zero contribution when evaluated between states whose charge
differs by one unit. Resolving the time-ordering and inserting a complete
set of states we find
\begin{eqnarray}
G_{Q}(x,x';t-t')& = &
\Theta(t'-t)\langle Q|\hat{\psi}_{sp}(x,t)|Q+1\rangle
\langle Q+1|\hat{\psi}^{*}_{sp}(x',t')|Q\rangle \nonumber \\
& + & \Theta(t-t')\langle Q|\hat{\psi}^{*}_{sp}(x',t')|Q-1\rangle\langle
Q-1|\hat{\psi}_{sp}(x,t)|Q\rangle.
\label{time}
\end{eqnarray}
The two terms in this expression correspond directly to the forward and
crossed channel processes shown in Figures 1 a) and 1 b) respectively.
The expectation values in (\ref{time}) can be evaluated using elementary
quantum mechanics; from (\ref{sp}) we find
\begin{equation}
\langle Q_{1}|\hat{\psi}_{sp}(x,t)|Q_{2}\rangle =\delta_{Q_{1}+1,Q_{2}}
\frac{1}{\lambda}
\frac{\sin\left(\frac{\lambda^{2}\bar{Q}}{4}\right)}{\cosh\left(m\sin\left
(\frac{\lambda^{2}\bar{Q}}{4}\right)x\right)}\exp\left[it(M(Q_{2})-M(Q_{1}))
\right],
\label{exp}
\end{equation}
where $\bar{Q}=(Q_{1}+Q_{2})/2$ and
$M(Q)=(4m/\lambda^{2})\sin(\lambda^{2}Q/4)$. The midpoint combination
$\bar{Q}$ arises because of the Weyl ordering of the field operator. All
that remains is to extract the the contribution to the T-matrix via the
LSZ reduction formula:
\begin{eqnarray}
{\cal T} & = & \int \frac{dx\,dt}{\sqrt{2\omega}}\exp(-ikx+iwt)
(\partial_{t}^{2}-\partial_{x}^{2}+m^{2}) \nonumber \\
& & \qquad{} \qquad{} \times \, \int \frac{dx'dt'}{\sqrt{2{\omega}'}}
\exp(ik'x'-i{\omega}'t')(\partial_{t'}^{2}-\partial_{x'}^{2}+m^{2})
G_{Q}(x,x';t-t'),
\label{tmatrix}
\end{eqnarray}
where $\omega^2=k^{2}+m^{2}$ and $\omega'^{2}=k'^{2}+m^{2}$. The most
important point of the analysis given in \cite{dor1} was to show that
$G_{Q}$ has poles in momentum space exactly on the mass shells of the
incoming and outgoing mesons and thus
yields a non-vanishing contribution to ${\cal T}$. To see this note
that the saddle point field $\psi_{sp}$ has the asymptotic behaviour
\begin{equation}
|\psi_{sp}(x,t)|\rightarrow \frac{2}{\lambda}\sin\left(\frac
{\lambda^{2}Q}{4}\right)\exp\left(-m\sin\left(\frac{\lambda^{2}Q}{4}\right)
|x|\right),
\label{asy}
\end{equation}
as $|x|\rightarrow\infty$. This means that we can
write the expectation value (\ref{exp}) as
\begin{equation}
\langle Q|\hat{\psi}_{sp}(x,t)|Q+1\rangle = \int\frac{d\omega\,
dk}{(2\pi)^{2}}\frac{{\cal A}(k)}
{k^{2}+m^{2}\sin^{2}\left(\frac{\lambda^{2}\bar{Q}}{4}\right)}\delta(\omega-
M(Q+1)+M(Q)),
\label{ft}
\end{equation}
where the residue ${\cal A}(k)$ remains finite at
$k^{2}=-m^{2}\sin^{2}(\lambda^{2}\bar{Q}/4)$. Hence the Fourier transform of
the meson Greens function has a pole at this point. The
meson mass shell condition is $k^{2}=m^{2}-\Delta M^{2}$ where $\Delta
M=M(Q+1)-M(Q)$. It is straightforward to check that this value of
$k^{2}$ coincides with
the position of the pole up to $\lambda^{4}$. Hence the momentum-space
Green's function $\tilde{G}_{Q}$ has a pole on the meson mass-shell
and yields a non-vanishing T-matrix element. The result is
\begin{equation}
{\cal T}_{1,Q}(\omega)=-\frac{16m^2}{\lambda^{2}}\left[\frac{\sin^{4}\left(
\frac{\lambda^{2}\left(Q+\frac{1}{2}\right)}{4}\right)}{\omega-M(Q+1)+M(Q)}
-\frac{\sin^{4}\left(\frac{\lambda^{2}\left(Q-\frac{1}{2}\right)}{4}\right)}
{\omega-M(Q)+M(Q-1)}\right].
\label{final2}
\end{equation}
As we show in Section 3, the resulting
contribution to the S-matrix has two simple poles whose positions and
residues match those of our conjectured scattering amplitude
at leading order.
Due to the cancellation of the leading order in $\lambda$ between the
two terms in the above expression,
this contribution to the scattering amplitude
is of $O(\lambda^{0})$. In fact, there is also another contribution to the
S-matrix at this order in the semiclassical approximation
which comes from the meson propagator in the soliton background
\cite{dor2}. By analogy with a similar analysis for the $\phi^{4}$
kink \cite{kaw} we expect that this term,
when added to the $O(\lambda^{0})$ part of
(\ref{final}), reproduces the naive leading-order result given in the
text, equation (\ref{smatrix1}).
\section*{Figure Captions}
\paragraph{}
Figure 1: Contributions of the states with charges $Q\pm 1$
to the meson-soliton scattering amplitude.
\paragraph{}
Figure 2: Contributions of the states with charges $Q_{1}\pm Q_{2}$
to the scattering amplitude for two solitons with charges
$Q_{1}$ and $Q_{2}$.
\paragraph{}
Figure 3: Contribution of two-particle intermediate states which
produce anomalous thresholds in the soliton-soliton scattering
amplitude.

\end{document}